# Web Standards as Standard Pieces in Robotics

## Sekou L. Remy[1]

**Abstract**

Modern robotics often involves the use of web technologies as a means to cope with the complexity of design and operation. Many of these technologies have been formalized into standards, which are often avoided by those in robotics and controls because of a sometimes warranted fear that "the web" is too slow, or too uncertain for meaningful control applications.

In this work we argue that while web technologies may not be applicable for all control, they should not be dismissed outright because they can provide critical help with system integration. Web technologies have also advanced significantly over the past decade. We present the details of an application of a web server to perform open- and close-loop control (between 3Hz and 1kHz) over a variety of different network topologies. In our study we also consider the impact of a web browser to implement the control of the plant. Our results confirm that meaningful control can be performed using web technologies, and also highlight design choices that can limit their applicability.

# 1 Introduction

Modern robots are being applied in a dizzying array of tasks to solve a number of complex challenges. Unfortunately, as the complexity of the tasks increases, and as the complexity of the domain of available tools also increases, the development/programming of the robots may be just as difficult. The resulting complexity in robotic development requires a plurality of functional expertise (e.g. control systems, system administration, software engineering, mechanical engineering, etc.). As such, a significant obstacle in robotic development are the system integration challenges associated with getting tools and quickly getting these tools to work together.

Service-Oriented Computing [7, 2] is a paradigm that leverages the creation of basic blocks to quickly develop reusable applications [12]. The service-oriented approach is interesting in robotics because it readily provides an approach that permits subject matter experts to encapsulate their knowledge into useful building blocks that can then easily be used by others. ROS [13] and other service-oriented robotic offerings like those published in [3] are making effective strides, but more is needed.

We believe one path to accelerate service-oriented development in robotics builds upon advances in Networked Control Systems. These are control systems where networks provide the connections between spatially distributed

S.L. Remy is with the School of Computing, Clemson University Clemson, SC 29634, USA `sremy at clemson.edu`

system components [8]. In this work we discuss how the web, and standards developed for it that permit it to leverage the network, can more broadly be applied in robotics.

We begin with a brief characterization of the spectrum of web standards that can be applied in robotics. We then select one configuration of components selected from this spectrum and exploit it to demonstrate both the limitations and the opportunities for robotics and control.

Although not formally recognized in the literature, we believe that one source of objections relating to incorporating "the web" into robotics is related to timing variability and delay. These properties have a significant impact on feedback control systems. Therefore, exposing the impact that web standards can have on these properties is a valuable research contribution.

## 2 Background

Combining robotics and web technologies is not a new concept. As early as the mid-1990's robotic arms and other devices were being shared on the web for public and private use [4, 1]. At that time, sensing and actuation were severely constrained by the data rates of the existing technology. As such, teleoperation was, at best, tolerable; and meaningful automatic control tasks were not realistic applications of the technology.

Since that time, data rates on the Internet have increased by three orders of magnitude [11]. Accompanying such developments in Internet and computing technologies also have been advances in sensing and actuation. These advances have been built upon processing and storage, which are both faster and consumes less power, and all this change provides opportunities for personal and industrial applications of web technologies – if they can be effectively harnessed in a principled manner. These opportunities can be advanced through the use of web standards.

### 2.1 Ecosystem of relevant standards

The Internet Engineering Task Force (IETF) and the World Wide Web Consortium (W3C) are two of the more visible bodies that are responsible for web (and Internet) standards. In fact, due to the omnipresence of the web in modern life, bodies like these increasingly function as standardization bodies for broader computing. For example, neither PNG (RFC2083) nor JPEG (RFC2045/2046) is data types that are just considered in web applications. They are common formats chosen for images in wherever computers are used. As another example, Bluetooth and Wi-Fi are two networking standards that are defined in the IEEE-802 suite of definitions. These standards are widely used in devices from televisions and printers, to the recently announced MYO, a Bluetooth connected EMG sensor.

In addition to data types and networking, there are protocols like HTTP (RFC 2616) that enable data communication over the multiple networks. Standardization has also enabled resources to be identified in a principled manner in this highly complex and dynamic environment. Such Uniform Resource Identifiers (URIs), defined in RFC3986, also include key=value pairs which readily enable data to be encoded for robotics and control use.

Finally, because these standards do not operate in a vacuum, the ecosystem around them is strongly influenced by their existence. For example, web servers and web browsers support many of these formal standards and although they are not formally sanctioned by any agency, they provide an informal standard which can be leveraged for our needs.

### 2.2 Approaches to leverage standards

We have presented the plurality of standards that are provided for the web, and in this section we describe how these standards can be combined in one configuration relevant to robotics, and subsequently we assess/quantify the characteristics of their use.

Specifically we leverage a web server, a RESTful web-service and clients written in python and javascript to show how standards that these components are based upon can be leveraged for autonomous control. Leveraging standards in this way provides a path to avoid the development of "yet another framework" that only runs on a given operating system, or with a given programming language, or only with a particular type of task. In this approach we demonstrate how operating frequency and networking conditions affect control. Our aim is to convince the reader that 1) there are many practical web standards that can be leveraged in robotics, 2) while "the web" may in general be the wild west of computing, there are configurations that can be effectively applied in relevant control.

# 3 Experiments

## 3.1 Configurations

In this work, we implement two types of clients, one written as a client-side javascript application and the other written in python. For both types of clients blocking requests are made to the server. This means that the client code halts until the server provides a response, or a time-out is triggered.

In the first portion of this work, the server's response is solely an acknowledgement that it has received data. This is a precursor to feedback control and is used to characterize the timing. Subsequently the server will respond with the updated state of the system coupled with it.

Usability studies [9, 10] have long suggested human computer users demand response times on the order of 100-250ms. While typical web browsing does not require constant refreshing of content, this response time suggests that the underlying Internet infrastructure can permit information to change at rate of 4 Hz. As such we will run each client at 3.3 Hz, 10 Hz, and 100 Hz. The 100 Hz number is selected to show the potential of these resources in significant control applications, like quadrotor attitude control [6] or control of a PUMA robotic arm [5].

### 3.1.1 Locations and Computers

The experiments in this work were performed with clients running on four networked computers. The first machine, *arima*, is also running the web server. *Arima* is wirelessly connected to the network using 802.11g. The second machine, *buccoo*, sits on a wired gigabit ethernet connection to the network. The third machine, *couva*, also uses a wireless 802.11g network connection, but unlike *arima*, it is connected to a WPA2 encrypted network. WPA2 (also called WPA-802.1X) is a networking standard that defines security for wireless networks and is associated with increased network overhead. The fourth machine, *diego*, is housed in a remote data center. This machine was accessed in two ways, resulting in different network topologies. The first involved a wired gigabit ethernet connection, while the second leveraged a Virtual Private Network.

Table 1: Relevant properties for computers used in testing.

| Client | Firefox | Python | Processor |
|---|---|---|---|
| arima | 6.02 | 2.6.6:84292 | 2.4 GHz Core 2 Duo |
| buccoo | 3.6.12 | 2.6.5:79063 | 2.4 GHz Core i7 |
| couva | 3.6.13 | 2.6.5:79063 | 1.66 GHz Atom N450 |
| diego | N/A | 2.7.1:86832 | 2.4 GHz QEMU Virtual |

All communication in this work leverages HTTP. As indicated by the ping statistics (Table 2), blocking communication between client and server can take up to seconds to occur, but characteristically occurs on the order of tens of milliseconds. Packet loss is a possibility in network communication; however, the TCP provides a reliable connection between nodes and will retransmit lost packages (invariably adding delay).

A five second timeout is used in the python clients, and the javascript clients in Firefox are also documented to by default also have a five second communication timeout that is known as an "unresponsive script" timeout.

Table 2: Statistical summary of the response packets received from ping commands between the server (*arima*) and each of the machines used to run the clients. diego[2] includes an off-campus, VPN facilitated communication.

| Client | **min**(ms) | **avg**(ms) | **max**(ms) | σ(ms) | loss |
|---|---|---|---|---|---|
| arima | 0.031 | 0.108 | 0.625 | 0.069 | 0.0 |
| buccoo | 1.360 | 3.376 | 4.539 | 0.484 | 0.0 |
| couva | 4.896 | 104.188 | 3033.374 | 336.043 | 1.6% |
| diego[1] | 0.943 | 3.155 | 13.42 | 0.752 | 0.9% |
| diego[2] | 39.01 | 53.14 | 145.2 | 21.75 | 0.0 |

### 3.1.2 Data Collected

In the message sent from client to server in each of the runs, we log the time that the message was sent and the time it was received. We do not perform clock synchronization between the two machines, so all reported results are based on relative time changes with the candidate reference points as the first noted time stamp for each of the time sources.

## 3.2 Closed-loop Control

To explore the impact of timing on closed-loop control, the client and the server are coupled with digital implementation of a motor controller and a D.C. motor. The transfer function for the controller and the plant are selected from [14]. In the original work these authors investigated the performance degradation introduced by delays in the control loop, and they are used for the same reason here.

The transfer function implemented in the client is defined as follows:

$$G_C(s) = \frac{a_0 s + a_1}{s}, \qquad (1)$$

where $a_0 = 0.1701$ and $a_1 = 0.378$. In the server, the plant's transfer function is:

$$G_P(s) = \frac{b_0}{(s+c_0)(s+c_1)}, \qquad (2)$$

and $b_0 = 2029.826$, $c_0 = 26.29$, and $c_1 = 2.296$.

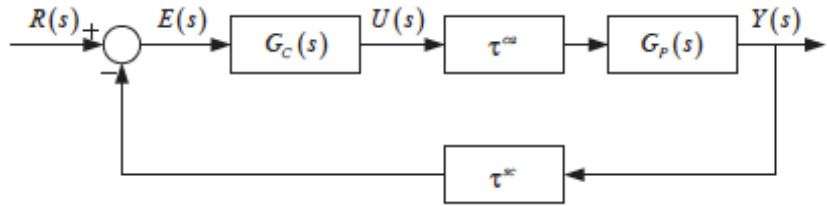

Figure 1: Closed-loop control system example.

This closed-loop control system, captured in Fig. 1, performs PI control on the plant, and is slightly different from the cited source. In this original work, $\tau^{ca}$ and $\tau^{sc}$ (the delay between controller and actuator, and between sensor and controller) are experimental parameters. In this work, however, these two delay values are influenced only by the location of the client. Another difference is that the values of $\tau^{ca}$ and $\tau^{sc}$ are only weakly correlated. While data often travels the same path to and from the server, it is not guaranteed to do so. Furthermore, even if the path remained consistent, the conditions along the way can vary from time to time. For this reason, the assumption that $\tau^{ca} = \tau^{sc}$ does not hold.

For both client and server, to discretize the transfer function, the backward Euler method is applied. This method, widely used in commercial control applications, is one integral based, numerical approximation method. It is applied by using the substitution formula in (3),

$$s = \frac{z-1}{zh}, \qquad (3)$$

where *h* is the simulation timestep. In this work, the same timestep is applied in both the controller and the plant. Values of *h=3ms* and *h=1ms* were both applied in this work.

# 4 Results

While TCP guarantees in-order and lossless communication between end points, there is no guarantee that the web server will process all messages in order. Due to factors such as packet buffering, queuing, message retransmission, and operating system scheduling, it is possible for sequential network packets to be provided to the server at the same time. When messages arrive at the server at the same time, the server spawns threads to process them. No guarantees can be made on which threads complete their processing first however. As such, it is possible for messages that arrive at the same time to be applied out of order in spite of TCP's sequential guarantees.

Table 3: Percentage of packets sent from javascript clients that arrived at the server (*arima*) at the same time as the previous packet in the sequence during 600 seconds of client operation.

| Client | 100Hz | 10Hz | 3.3Hz |
|---|---|---|---|
| arima | 0.2494 | 0.0 | 0.0 |
| buccoo | 0.0746 | 0.0 | 0.0010 |
| couva | 0.0882 | 0.0051 | 0.0005 |

As indicated in Table 3, this occurrence is affected by both the operating frequency of the client and the network connection to the server. The effect of occurrences like these is that control commands can be issued by the server out of order. This means that control noise is introduced in the data generated by the client.

While this effect is magnified by higher operating frequencies, its impact is also marginalized with faster control. Since higher operating frequencies result in shorter time between control actions, there is also less time for the system dynamics to develop. In this work, this effect is noticed as additional noise in the inter-departure times and the existence of zero values in the inter-arrival times. Inter-arrival times should not be less than the inter-departure times, which in the ideal case are the inverse of the operating frequency.

## 4.1 Timing with 100 Hz Javascript Client

In figure 2 we show the first 600 samples of the time series generated by the javascript client running on *arima*. This figure captures variability, delay and a dilation effect (the altered slope) that is likely to be related to processing overhead.

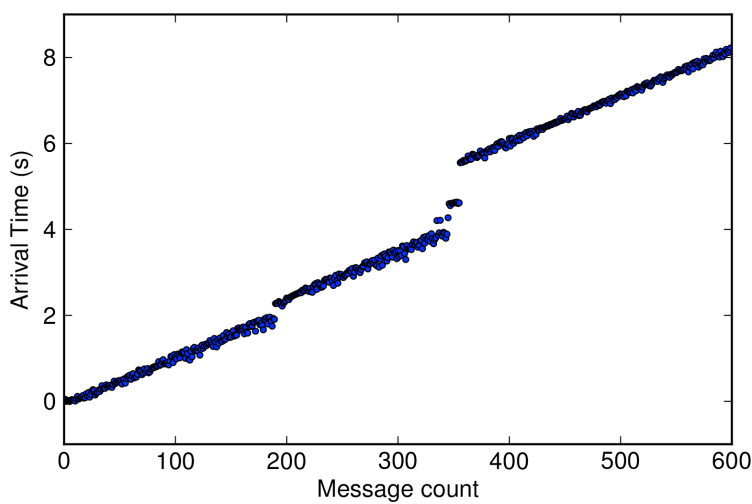

Figure 2: First 600 arrival times for messages from the javascript client running at 100 Hz on *arima*.

The last two of these terms, overhead and transmission delay, both have long-term impact on the number of messages that are transmitted during control. Specifically, they both result in increased system latency. The 600th message should have been sent six seconds after the first message. This graph shows that it was sent two seconds late. In the first six seconds, just shy of 400 messages were transmitted. Over the ten minutes of the client's operation, only 55 thousand messages were transmitted, better than the effective rate observed during the first six seconds, but still short of the 60 thousand messages that should have been sent.

If open-loop control was performed under such conditions, this would mean that after 600 seconds of operation, the controller would be issuing control that is roughly 50 seconds delayed. For every ten seconds of control, the controller falls an additional second behind. For applications with long standing interaction, this would be quite problematic. When closed-loop control is considered, however, this is diminished. After ten minutes of operation, the controller would still be operating on the most recently received input, which according to the data from traceroute and ping commands (See Table 2), should only be delayed on the order of tens of milliseconds.

## 4.2 Impact of client location

To explore how the timing vagaries are affected by the selection of different clients, we next consider how the messages generated by the clients compare to the desired reference timeline. We test this on each of the three machines and first consider javascript and python clients running at 3.3Hz.

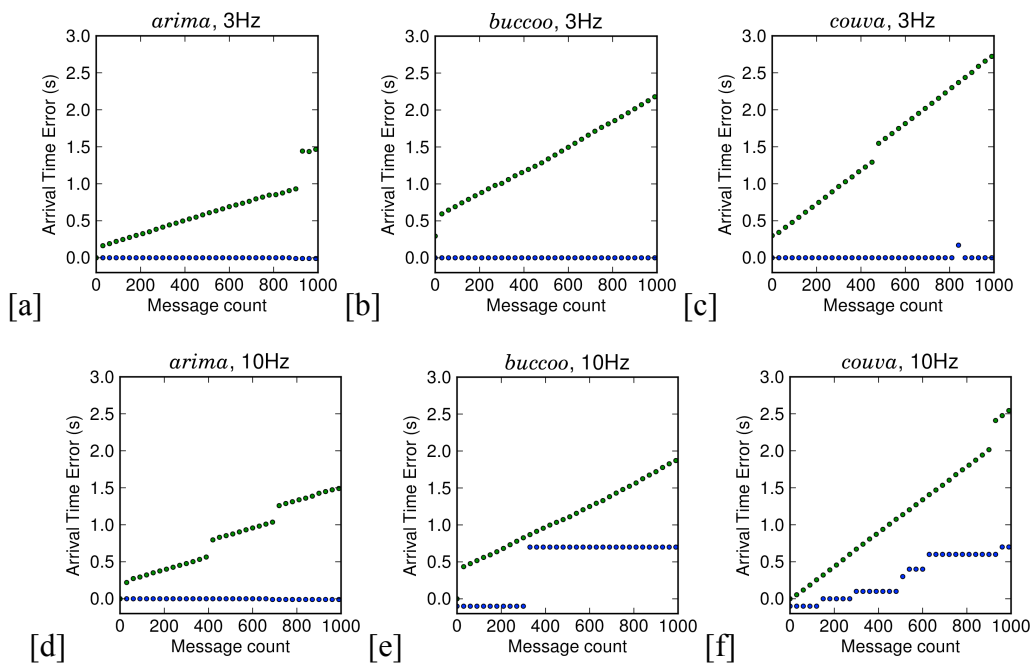

Figure 3: The difference in the time stamps of generated messages from python (blue) and javascript (green) operating at 3.3Hz on *arima* (3a), on *buccoo* (3b), and on *couva* (3c). Figures 3d-3f are for 10Hz control on these machines as well.

In figure 3, the plots of error for the respective clients are shown. In all of these cases, the python client effectively generates messages that are received on the desired timeline. This is evident because there is little difference between the timestamps for the client messages and the ideal values (zero slope). What should also be evident from these figures is that the javascript clients all diverge from the reference with time (non-zero slope).

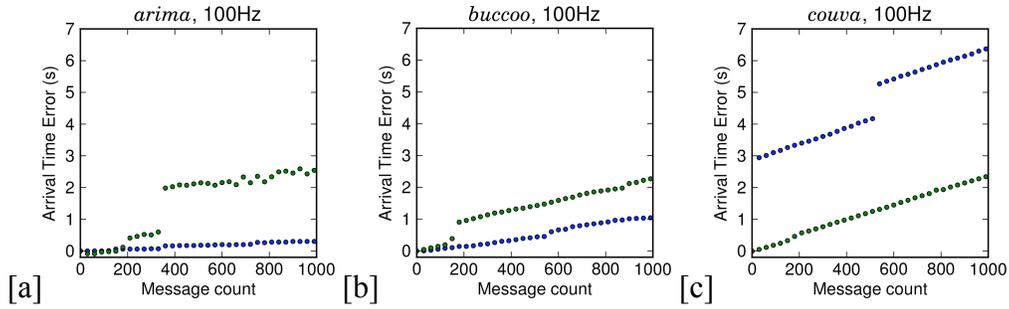

Figure 4: The difference in the time stamps of generated messages from python (blue) and javascript (green) operating at 100Hz clients on each of the three machines.

The same basic trend is observed at higher operating frequencies. At these higher frequencies, the impacts of the network delay and network jitter are more pronounced because the messages are being transmitted more frequently. At the same time, the impact of noise in the client timing infrastructure is also more pronounced. One millisecond of error is more significant in ten millisecond intervals than it would be when that interval is 100 or 300 milliseconds. Messages should have the same base transmission delay, but the effects of queuing and conflict resolution for the shared medium are now amplified. (Recall that the server's acknowledgement is also guaranteed to be sent back to the client.) These effects are evident in the non-zero differences from the reference.

At 10Hz, even though the python client (See Figures 3d-3f has non-zero differences from the reference, the deviations are largely isolated to impulses (i.e. errors injected at instances) not the continuous deviation that appears constant for javascript clients. For these clients, there is no significant divergence as the slopes of data (in blue) are essentially zero. As the highest operating frequency, 100Hz, such stability no longer exists in the data (See Figures 4a-4c).

## 4.3 Closed-loop Control

From the figures presented thus far, it is clear that undesirable timing characteristics exist between the clients and the server. To contextualize the impact of these, closed-loop control was performed over the network and facilitated by the web server, URI encoded queries and a web browser to run javascript versions of the client. The operating frequencies in these closed-loop cases are higher than the 100Hz previously considered to enable comparison with published results.

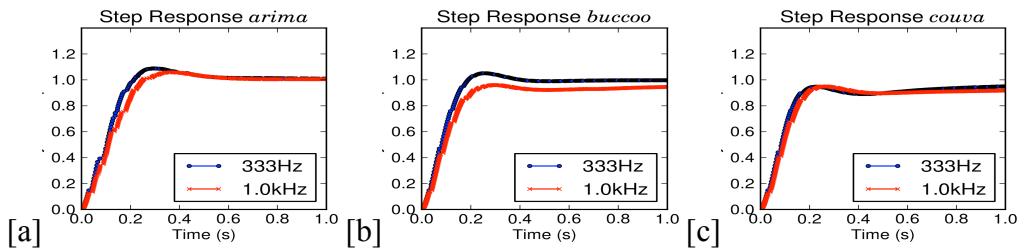

Figure 5: Unit step responses for controllers and plants running under different network locations.

As the graphs in Figure 5 confirm, even in the light of these timing challenges, closed-loop control with python controllers on *arima*, *buccoo*, and *couva* were all successfully demonstrated.

The performance is in keeping with [14] and even when the operating frequency is increased to 1 kHz, effective control was performed. Recall that this controller was not designed for network use, and makes no accommodation for the timing challenges of distributed control. When the javascript controller is used however, the settling time of the system increases ten-fold (see Figure 6a). The system remains stable, but would not be practical for use where response times on the order of fractions of seconds are required. It may be possible to leverage Kalman filters to avoid this issue, but it is not yet clear what circumstances should warrant such efforts.

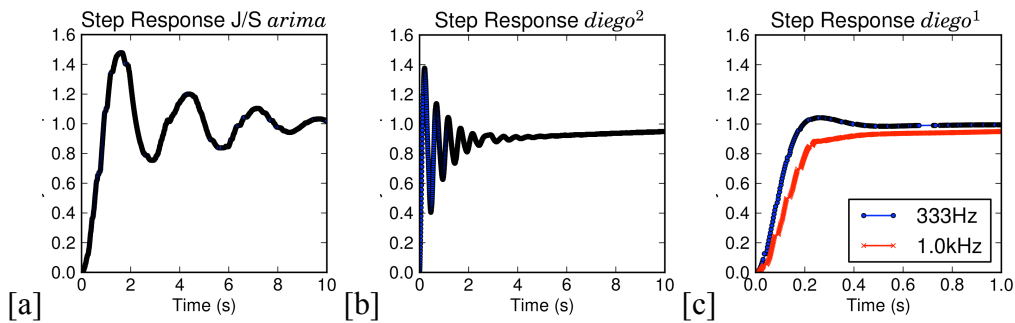

Figure 6: Unit step responses operating at 333Hz for javascript client on *arima* (6a), and for python clients on *diego* connected via a VPN from off-campus (6b) and over the campus LAN (6c).

Surprisingly, as seen in Figure 6b, when a python controller is run from an on-campus location, to an off-campus plant (server), the settling time was only quadrupled. If that same location is used and the plant is moved on-campus, its operation is in keeping with performance on *arima*, *buccoo*, and *couva* for both 333Hz and 1KHz control.

Plainly put, these results confirm that meaningful closed-loop control, in this case control of D.C. motor with time constant 30*ms*, is only negatively impacted the use of web standards when a javascript client is used. The javascript client, which runs in-browser, is crippled because it shares the same thread for control and the other core functionality of the browser's operation. The usefulness of this resource for high-speed control is thus significantly hindered by this operating constraint.

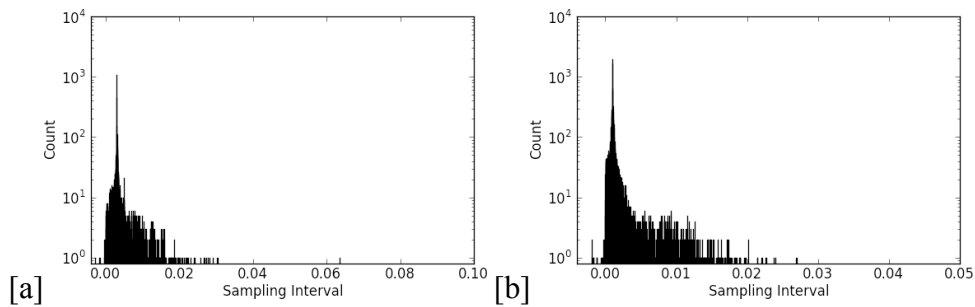

Figure 7: Sample intervals observed for the server operating with 0.001s intervals and 0.003s intervals (Figure 7a and 7b respectively).

To reveal why the system is able to successfully operate under these circumstances, we explore how the intervals in which new control signals are generated are effected by client/server locations.

As can be seen in Figure 7, the sampling intervals on the computers are quite noisy. The higher the frequency, the larger the variation in the intervals. As expected, the means of these intervals are 3.0*ms* and 1.004*ms* while their standard deviations are 2.984*ms* and 3.705*ms* respectively.

It is known that real-time operating systems have merit in control, but these plots highlight precisely how. Even before considering the variations introduced by the use of the network or web standards, there is noise injected into system control. The reduction in the number of negative and zero-length intervals also evident in this figure is associated with the additional overhead to calculate control values, a factor not previously present for the data in Table 3.

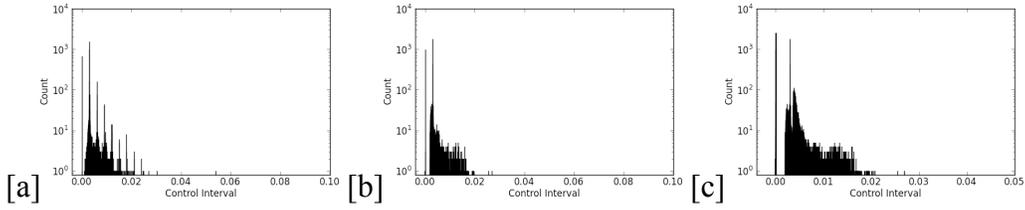

Figure 8: Distribution of control intervals observed for controllers on Arima (a), Buccoo (b), and Couva (c) operating at 333Hz (each should be 0.003s).

The impact of variability can also be seen when comparing the histograms in Figure 8. For the client that is closest to the server (Figure 8), the impact of the controller sampling frequency is the dominant factor influencing the control intervals. In this case, since the sample time in on the order of the transmission time, it is possible, especially in the face of the sampling noise, to receive multiple control signals during a single sampling interval. It is also possible to send multiple control signals before the response is received from the plant (and new control values are generated). These two factors account for the strong presence of periodic intervals that are multiples of the sampling time, and also for zero-length intervals.

As the transmission time for control signals increased (Figures 8b and 8c), the likelihood of sending multiple control signals before receiving a response increases. At the same time, the possibility of the plant receiving multiple control signals in a sample interval decreases. These two changes are appropriately reflected in these histograms (flatter noise profile, and more zero-length intervals), as well as in Figure 9b.

These graphs all suggest that the use of a desktop operating system, and communication with very high transmission latency have a more significant impact than the use of the web server, and URI encoded control (and responses).

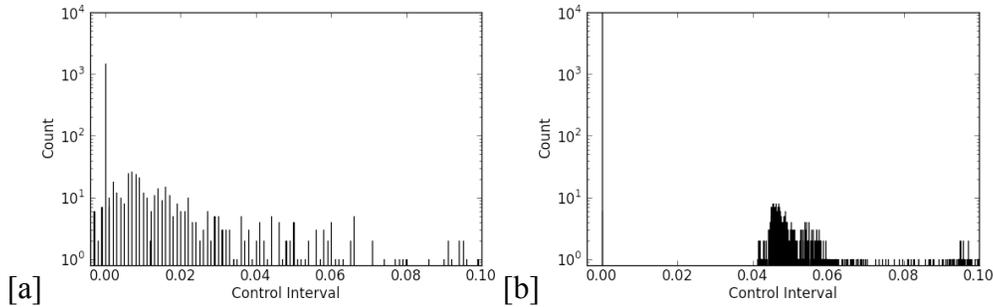

Figure 9: Undesirable effects of client type and network delay on the desired control interval (0.003s). Figure 9a is was measured for the javascript client running on *arima* 9b is for the python client running on *diego* connected via a VPN from off-campus.

The timing challenges of using javascript at this frequency has already been stated. Neither the browser, nor the scripting language was designed to operate in this manner, and it is clear when one considers the wide distribution of control intervals associated with 3*ms* intervals (See 9a). Design notwithstanding, this histogram shows support for avoiding control with time intervals on the order fractions of seconds. This guideline should be reconsidered within 2 years, as modern browsers are transitioning to different interaction paradigms and implmentations.

Considering the histogram for control intervals to an off-campus controller (Fig. 9b), the impact of transmission delay at minimum on the order of 40*ms* is evident. As previously indicated, the increased transmission time results in more zero-length intervals and although not even the target sampling rate is evident, control is still able to be stably demonstrated in the system.

# 5 Conclusion

From these results we can conclude that relevant control can be performed at operating frequencies meaningful in robotics. Further study is needed with systems that are unstable to characterize the opportunities for other relevant classes of robotic systems.

This work shows that by leveraging resources and standards long pervasive in networked control environments, that exciting opportunities to connect to high-level distributed control systems, with little additional investment by research-ers and developers. Web standards and browser standards are powerful, yet underutilized tools in many computing applications. To enable these resources to be more widely appreciated, we advocate that meaningful collaboration must occur across disciplines and many practical activities attempted to permit researchers to learn each other's languages and to recognize each other's needs. In this way our fields can become better equipped to address the many complex and interesting challenges that we have long dreamed to address.

# References

[1] P. Backes, G. K. Tharp, and K. S. Tso. The web interface for telescience (wits). In *in Proc. IEEE Int. Conf. Robot. Automat*, pages 411–417, 1997.
[2] M. Blake, S. Remy, Y. Wei, and A. Howard. Robots on the web. *Robotics Automation Magazine, IEEE*, 18(2):33 –43, june 2011.
[3] Y. Chen, Z. Du, and M. Garcia-Acosta. Robot as a service in cloud computing. In *Service Oriented System Engineering (SOSE), 2010 Fifth IEEE International Symposium on*, pages 151 –158, june 2010.
[4] K. Y. Goldberg, M. Mascha, S. Gentner, N. Rothenberg, C. Sutter, and J. Wiegley. Destop teleoperation via the world wide web. In *ICRA*, pages 654–659, 1995.
[5] S. Hayati, T. Lee, K. Tso, P. Backes, and J. Lloyd. A unified teleoperated-autonomous dual-arm robotic system. *Control Systems, IEEE*, 11(2):3–8, Feb. 1991.
[6] G. M. Hoffmann, S. L. Wasl, and C. J. Tomlin. Quadrotor helicopter trajectory tracking control. In *In Proc. AIAA Guidance, Navigation, and Control Conf*, Aug. 2008.
[7] M. Huhns and M. Singh. Service-oriented computing: key concepts and principles. *Internet Computing, IEEE*, 9(1):75 – 81, jan-feb 2005.
[8] H. Lin, G. Zhai, and P. Antsaklis. Robust stability and disturbance attenuation analysis of a class of networked control systems. In *42nd IEEE Conference on Decision and Control*, volume 2, pages 1182 – 1187, Dec. 2003.
[9] R. B. Miller. Response time in man-computer conversational transactions. In *Proceedings of the December 9-11, 1968, fall joint computer conference, part I*, AFIPS '68 (Fall, part I), pages 267–277, New York, NY, USA, 1968. ACM.
[10] J. Nielsen. *Usability Engineering*. Morgan Kaufmann Publishers Inc., San Francisco, CA, USA, 1993.
[11] J. Nielsen. Nielsen's Law of Internet Bandwidth. *http://www.nngroup.com/articles/nielsens-law-of-internet-bandwidth/*, Apr. 1998. Accessed 2013-01-02.
[12] M. P. Papazoglou, P. Traverso, S. Dustdar, and F. Leymann. Service-oriented computing: a research roadmap. *International Journal of Cooperative Information Systems*, 17(02):223, 2008.
[13] M. Quigley, K. Conley, B. P. Gerkey, J. Faust, T. Foote, J. Leibs, R. Wheeler, and A. Y. Ng. ROS: An Open-Source Robot Operating System. In *ICRA Workshop on Open Source Software*, 2009.
[14] Y. Tipsuwan and M.-Y. Chow. Control methodologies in networked control systems. *Control Engineering Practice*, 11(10):1099–1111, 2003.